 \documentclass[preprint,3p,times,sort&compress]{elsarticle}

\usepackage{amssymb}
\usepackage{hyperref}
\usepackage{epstopdf}

\journal{Journal of Magnetism and Magnetic Materials}

	\usepackage{fancyhdr}
\renewcommand{\thispagestyle}[1]{} 

\begin{document}

\begin{frontmatter}



\title{Graphene nanoflakes in external electric and magnetic in-plane fields}


\author{Karol Sza{\l}owski}

\ead{kszalowski@uni.lodz.pl; kszalowski@wp.pl}
\address{Department of Solid State Physics, Faculty of Physics and Applied Informatics, University of {\L}\'{o}d\'{z}, ul. Pomorska 149/153, 90-236 {\L}\'{o}d\'{z}, Poland}

\date{\today}

\begin{abstract}
The paper discusses the influence of the external in-plane electric and magnetic field on the ground state spin phase diagram of selected monolayer graphene nanostructures. The calculations are performed for triangular graphene nanoflakes with armchair edges as well as for short pieces of armchair graphene nanoribbons with zigzag terminations. The Mean Field Approximation (MFA) is employed to solve the Hubbard model. The total spin for both classes of nanostructures is discussed as a function of external fields for various structure sizes, for charge neutrality conditions as well as for weak charge doping. The variety of nonzero spin states is found and their stability ranges are determined. For some structures, the presence of antiferromagnetic orderings is predicted within the zero-spin phase. The process of magnetization of nanoflakes with magnetic field at constant electric field is also investigated, showing opposite effect of electric field at low and at high magnetic fields. 

\end{abstract}


\end{frontmatter}

\section{Introduction}

Graphene, being the first truly two-dimensional material, exhibits highly unusual electronic properties, and offers, in addition to novel, unique physics, also a promising platform for applications in spin electronics \cite{Bullard2015,Han2014,Roche2014,Rycerz2007}. As a consequence, intense studies of magnetic properties of graphene are highly encouraged \cite{Yazyev2010}. Within this field, recent experimental findings concerning the existence of magnetically polarized edge state in graphene nanoribbons \cite{Magda2014} provide a sound motivation for theoretical studies of magnetism in geometrically constraint graphene. The presence of edge and further reduction of dimensionality of graphene below two significantly modifies its electronic structure, causing the emergence of early-predicted \cite{Wakabayashi1996} and experimentally observed \cite{Klusek2000,Klusek2005,Kobayashi2005} edge states. On the other hand, it also enables the design of the energy spectrum \cite{Yazyev2013}, which crucially shapes the magnetic behaviour.

Magnetism arising in zero-dimensional graphene-based nanostructures (nanoflakes, quantum dots, nanoribbons) as a result of the presence of edge and modification of electronic structure has recently collected rich literature \cite{Hikihara2003,Rossier2007,Ezawa2007,Wang2008,Wang2008b,Wang2009,Guclu2009,Feldner2010,Feldner2011,Potasz2012,Ezawa2012,Guo2013,Golor2013,Golor2013b,Ominato2013,Ominato2013b,Golor2014,Kabir2014,PotaszBook,Chacko2014,Cheng2014}. The potential for spintronics applications has been proven by the proposals of numerous graphene nanoflake-based devices (e.g. \cite{Ezawa2009,Spin2013,Zhang2014}). In particular, the description of spin-polarized transport through graphene nanostructures was already a goal of numerous studies \cite{Krompiewski2009,Lipinski2010,Ma2011,Weymann2012,Weymann2012b,Krompiewski2012,Luo2014,Modarresi2014b}. This supports hopes for development of carbon-based spintronics.

The key feature for spintronics is the ability to control the magnetic properties. Therefore, one of the flourishing fields of exploration is the influence of external fields on the magnetic properties of graphene and nanographenes. Within this scope of studies, the electric field focuses dominant attention, both in the context of graphene itself \cite{Novoselov1,Killi2011,Santos2013a,Santos2013b,Parhizgar2013,Chen2013,Yun2014,Pike2014}, its cousin systems \cite{Kou2012,Dolui2012,Ouyang2014,Xiao2014} as well as for derivative nanostructures, like monolayer and bilayer graphene nanoflakes (quantum dots) \cite{Agapito2010,Guclu2011,Ma2012,Lu2012,Potasz2012,Zhou2013,Modarresi2014,Sheng2014,Szalowski2014,Farghadan2014,Zhou2014,Dong2014,Beljakov2014}, nanoribbons \cite{Son2006,Guo2008,Nomura2010,Rozhkov2011,Bundesmann2013,Yamanaka2014,Bao2014,Zhang2014,Derakhshan2014,Anka} or nanotubes \cite{Yamanaka2012}. The effect of magnetic field on magnetism in graphene-based structures has been also investigated \cite{Jaskolski2008,Rycerz2010,Guclu2013b,Szalowski2013c,Golor2013b,Droth2014}. However, it is rather rare to study the common influence of both fields on the properties of graphene \cite{Roslyak2010,Giavaras2012}. 

In order to extend our previous theoretical works concerning the sole effect of electric field \cite{Szalowski2014} or magnetic field \cite{Szalowski2013c} on selected aspects of magnetism in some graphene nanoflakes, we perform the present study. For example, Ref.~\cite{Szalowski2014} was focused strongly on indirect coupling between external magnetic planes mediated by a graphene nanostructure and mainly on the possibility of controlling it with electric field. Therefore, our present aim is to investigate the combined effect of external in-plane electric and magnetic field and to construct the ground state spin phase diagram of some monolayer graphene nanostructures. The calculations presented in the paper are based on the tight binding model with Hubbard term (in MFA), supplemented with electrostatic potential term and Zeeman term. 

Our systems of interest belong to a class of very small graphene quantum dots, so that the strategy to obtain them experimentally is likely a bottom-top approach, i.e. a procedure based on some molecular precursors. In such context, it should be strongly emphasized that some of the nanographenes with shapes and sizes corresponding to our interest in the present work have been recently synthesized within such scheme, what paved the way towards characterization of their magnetic properties \cite{Konishi2013,Konishi2013b} and served as a motivation for the present research. The experimental development of the class of carbon-based nanostructures under discussion would allow to extend the range of systems in which the control over magnetism by electric field was proven in experiment beyond 'classical' semiconducting systems (see e.g. \cite{Ohno2000,Sawicki2010}).

The system of interest as well as theoretical model and numerical results will be discussed in details in the following sections of the paper.

\section{Theoretical model}

\begin{figure}[h!]
  \begin{center}
   \includegraphics[scale=0.5]{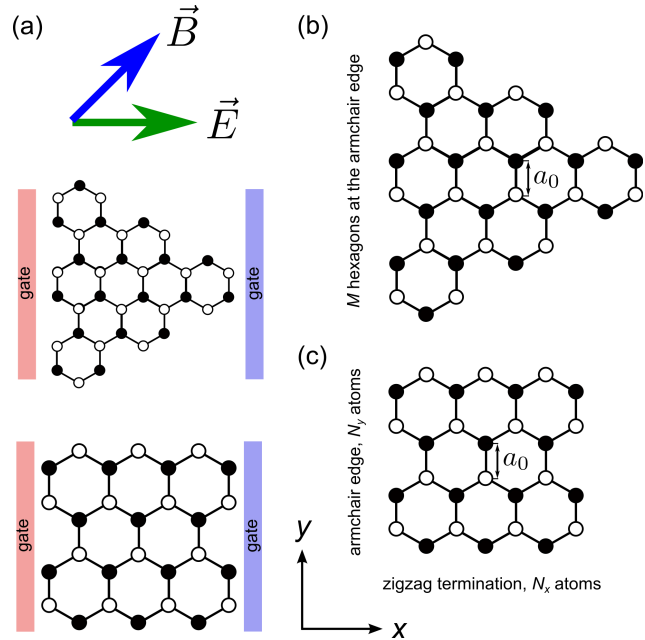}   
  \end{center}
   \caption{\label{fig:fig1}(a) Schematic view of graphene nanoflakes in external in-plane electric and magnetic field; (b) Schematic view of triangular graphene nanoflake with armchair edge composed of $M$ hexagons; (c) Schematic view of a short section of armchair graphene nanoribbon with zigzag terminations, width equal to $N_x$ and length equal to $N_y$.}
\end{figure}

The system of interest in the present paper is a monolayer graphene nanostructure (nanoflake or quantum dot) in external in-plane electric and magnetic field. The schematic view of the studied structures is shown in Fig.~\ref{fig:fig1}(a), where the orientation of nanoflakes with respect to the electric field (provided by the gates) along the direction $x$  is sketched. Also the example magnetic field direction is shown (however, it can be an arbitrary other in-plane direction). Each nanostructure is composed of $N$ carbon atoms, belonging to two interpenetrating sublattices (marked with filled and empty symbols). Our interest is focused on two classes of graphene nanoflakes (GNFs). The first one is a triangular nanoflake with armchair edges, each one consisting of $M$ hexagons, what defines the size of the structure (see Fig.~\ref{fig:fig1}(b)). The second class of structures is an ultrashort section of armchair nanoribbon with zigzag terminations (shown in Fig.~\ref{fig:fig1}(c)). This nanoflake is characterized by $N_x$ atoms constituting a zigzag termination and $N_{y}$ atoms along each armchair edge. We assume that all the edge carbon atoms are passivated.

We have selected such shapes of graphene nanostructures in general to avoid studying the commonly discussed case of spin-polarized zigzag edge state, for example in zigzag-edged triangular nanoflakes \cite{Potasz2012} or in zigzag-edged nanoribbons \cite{Son2006,Yazyev2010}. We would rather concentrate on the structures possessing armchair edges to avoid zero-field magnetic polarization.  Moreover, the present calculations \cite{Yamanaka2014} indicate that for some shapes and orientations of the nanostructures the screening effects for the electric field are particularly weak. The same factor limits our interests to nanostructures of small sizes (see also the discussion in our work \cite{Szalowski2014}). We should emphasize here that in order to avoid metallic character of our structures, we consider the smallest, molecular-like nanoflakes with well separated energy states and with a significant gap between highest occupied state and lowest unoccupied electronic state.

The goal of the model we use is to capture magnetic properties which result from the behaviour of $p^{z}$ electrons in graphene nanostructures. These electrons contribute the energy states close to the Fermi level for charge-neutral nanoflakes. In the present work we describe the behaviour of the mentioned charge carriers by means of the following tight-binding based Hamiltonian in real space:\cite{Szalowski2013c,Szalowski2014}
\begin{eqnarray}
\label{eq:eq1}
&&\!\!\!\!\!\!\!\!\!\!\!\mathcal{H}=-\sum_{<i,j>,\sigma}^{}{t_{ij}\,\left(c^{\dagger}_{i,\sigma}c_{j,\sigma}+c^{\dagger}_{j,\sigma}c_{i,\sigma}\right)}\nonumber\\&&\!\!\!\!\!\!\!\!\!\!\!+U\sum_{i}^{}{\left(\left\langle n_{i,\uparrow}\right\rangle n_{i,\downarrow}+\left\langle n_{i,\downarrow}\right\rangle n_{i,\uparrow}\right)}-U\sum_{i}^{}{\left\langle n_{i,\uparrow}\right\rangle\left\langle n_{i,\downarrow}\right\rangle}\nonumber\\&&\!\!\!\!\!\!\!\!\!\!\!+eE\sum_{i,\sigma}^{}{x_i\,n_{i,\sigma}}+\frac{\Delta_{B}}{4}\sum_{i}^{}{\left(n_{i,\uparrow}-n_{i,\downarrow}\right)}.
\end{eqnarray}
In the first term, $t_{ij}$ is the hopping integral between nearest-neighbour carbon sites. We assume that $t_{ij}=t$ for all nearest-neighbour bonds in triangular graphene nanoflakes with armchair edges. On the contrary, for ultrashort pieces of graphene nanoribbons with armchair side edges and zigzag terminations, we assume that $t_{ij}=t(1+\Delta)$ with $\Delta=0.12$ for outermost bonds at the armchair edges\cite{Son2006,Szalowski2014}. For the remaining nearest-neighbour pairs we take $t_{ij}=t$. The value of $t$ is usually set as 2.7 eV \cite{rmp}. The value of $t$ can serve as an useful energy scale in further considerations. The operators $c^{\dagger}_{i,\sigma}$ ($c_{i,\sigma}$) create (annihilate) an electron of spin $\sigma=\uparrow,\downarrow$ at lattice site $i$, while $n_{i,\sigma}=c^{\dagger}_{i,\sigma}c_{i,\sigma}$ is the number of such electrons. 

The tight-binding term is supplemented with Hubbard term, which is considered within MFA, thus neglecting the products of fluctuations $\left(n_{i,\uparrow}-\left\langle n_{i,\uparrow}\right\rangle\right)\left(n_{i,\downarrow}-\left\langle n_{i,\downarrow}\right\rangle\right)$. The on-site effective energy is taken as $U/t=1.0$ (see \cite{Schuler2013}, where such an assumption is justified for both charge-neutral and weakly doped structures). It should be emphasized that graphene and graphene nanoflakes do not belong to strongly correlated systems.

In order to account for the external in-plane electric field $E$ along $x$ direction, the electrostatic potential term is added. Moreover, the external in-plane magnetic field $H$ is introduced by means of Zeeman term, where $\Delta_B=g_{e}\mu_{\rm B}H$ is the Zeeman energy parameter. Let us mention that the in-plane orientation of magnetic field does not lead to the Peierls substitution\cite{Hofstadter}, because the hopping integrals are only modified in presence of in-plane component of the vector potential of the magnetic field. 

The horizontal electric field is assumed to originate from the gates between which the graphene nanostructure can be placed. In our considerations we use the normalized electric field, $Ea_0/t$. It might be useful to mention that for graphene nanostructures, the unity value of this normalized field would correspond roughly to 2 V/\AA. Regarding the magnetic field, let us mention that the source of Zeeman term present in our model may be either external magnetic field or, even more likely, it can result from exchange bias originating from the magnetic substrate on which graphene nanostructures can be deposited. Let us focus on the second possibility, being particularly interesting in the light of recent experimental achievements concerning growth of graphene on magnetic insulator EuO or yttrium iron garnet (see. e.g. \cite{Swartz2013,Swartz2013b,Tang2015}) as well as corresponding theoretical works about substrates such as EuO and BiFeO$_3$\cite{Yang2013,Qiao2014}. Let us mention that Ref.~\cite{Qiao2014} predicts the exchange field to be tunable via external pressure varying the distance between graphene and substrate and the calculated values of exchange splitting reach 150 meV, what corresponds to $\Delta_{B}/t$ about 0.05. Following the discussion regarding the exchange splitting in Ref.~\cite{Khodas2009}, it can be estimated that placement of graphene on magnetic substrates with even higher critical temperature that europium oxide would increase significantly the exchange splitting. Another scenario can be connected with inducing the splitting with magnetic impurities (e.g. \cite{Maria2010,Rappaport2011,Sung2014} or functionalization with molecular magnets \cite{Candini}. Such possibilities extend significantly the potential interest in the presented phase diagram by including substrate or magnetic proximity effects.

Let us now discuss some features of mean field approximation to the Hubbard model which we use in our work. Such an approach has a quite rich record in the recent literature on graphene magnetism (see e.g. \cite{Wakabayashi1996,Yazyev2010,Feldner2010,Szalowski2011,Szalowski2013a,SzalowskiAPPA,Szalowski2013c,Jaskolski2014,Magda2014,Bullard2015}). First we mention that this approximation breaks the rotational symmetry in spin space which is present in the original Hubbard term ($U\sum_{i=1}^{N}n_{i,\uparrow}n_{i,\downarrow}$). As a consequence, the total spin does not commute with the MFA Hamiltonian and is no longer a good quantum number. However, $z$-component of the total spin, $S=\sum_{i=1}^{N}(n_{i,\uparrow}-n_{i,\downarrow})/2$ is well defined for the eigenstates of MFA Hamiltonian. Therefore, in all further considerations, by total spin we will mean the $z$ component of it. 

Secondly, we may discuss briefly the MFA against other methods of solving the Hubbard model for graphene nanostructures. In comparison with Quantum Monte Carlo (QMC) and exact diagonalization performed in Ref.~\cite{Feldner2010}, it has been found that MFA provides a good tool to characterize the properties provided that $U/t$ is not excessively large. In particular, the most important quantity from our point of view is the total ground state energy, the knowledge of which is crucial for the construction of ground state phase diagrams. The comparison of this energy obtained within MFA and by exact diagonalization (smaller quantum dot systems) and QMC (larger quantum dot systems) yields the very good consistency for $U/t\lesssim 1$. Another quantity is the total $z$ component of spin (or staggered magnetization for antiferromagnetic orderings). This quantity calculated within MFA was also very close to the predictions of exact diagonalization and even closer for larger system solved by QMC for $U/t\lesssim 2$. Therefore, Ref.~\cite{Feldner2010} supports the applicability of MFA to the present model for the choice of $U/t=1$. Also in Ref.~\cite{Feldner2011} MFA has been discussed vs. QMC results for zigzag nanoribbon systems with the result that it provides a quantitatively reliable picture for nanoribbons wide enough. Also the sign of magnetic correlations is reproduced correctly (however MFA tends to overestimate somehow the range of correlations and may predict excessively robust magnetic orderings \cite{Feldner2011,Golor2014}).

 We believe that MFA provides a good reference point for studies of phase diagrams of graphene nanostructures of various size and shape, not being severely limited by rapidly growing computational demands, thus applicable to considerably large systems described by real-space Hamiltonians without translational symmetry. Also let us mention its favourable comparison with recent experiment \cite{Magda2014}.

The Hamiltonian (Eq.~\ref{eq:eq1}) can be decomposed into the form of $\mathcal{H}=\mathcal{H}_{\uparrow}+\mathcal{H}_{\downarrow}-U\sum_{i}^{}{\left\langle n_{i,\uparrow}\right\rangle\left\langle n_{i,\downarrow}\right\rangle}$. Then the pair of Hamiltonians $\mathcal{H}_{\uparrow}$, $\mathcal{H}_{\downarrow}$ can be subject to simultaneous, self-consistent numerical diagonalization starting from random initial conditions for $\left\langle n_{i,\sigma}\right\rangle$ until the convergence of eigenvalues $\epsilon^{\sigma}_{i}$ as well as convergence of charge densities $\left\langle n_{i,\sigma}\right\rangle$ is reached\cite{Szalowski2011,Szalowski2013c,Szalowski2014}. In the present work we used LAPACK \cite{lapack} package for this purpose. The diagonalization was performed in presence of constant total number $N_e$ of charge carriers. The situation of $N_{e}=N$ (one $p^{z}$ electron per carbon atom) corresponds to charge neutrality of the nanostructure, while $N_{e}=N+\Delta q$ describes doping with electrons ($\Delta q>0$) or holes ($\Delta q<0$). The diagonalization procedure was repeated many times to verify whether the true ground state, characterized by the minimal total energy of the charge carriers $\varepsilon_{total}=\sum_{i=k}^{N_{e}^{\uparrow}}{\epsilon^{\uparrow}_{k}}+\sum_{k=1}^{N_{e}^{\downarrow}}{\epsilon^{\downarrow}_{k}}-U\sum_{i}^{}{\left\langle n_{i,\uparrow}\right\rangle\left\langle n_{i,\downarrow}\right\rangle}$, was reached. Moreover, all the $N_{e}^{\uparrow}$ and $N_{e}^{\downarrow}$ values adding up to $N_{e}$ were tested to find the configuration minimizing the energy. The performed ground-state calculations correspond to $T=0$. As a result of the numerical procedure, $z$ component of the total spin as well as $z$ components of spin densities $s_{i}=\left(n_{i,\uparrow}-n_{i,\downarrow}\right)/2$ were obtained. This allowed to construct  ground-state phase diagrams in electric and magnetic field.

\begin{figure}[h!]
  \begin{center}
   \includegraphics[scale=0.25]{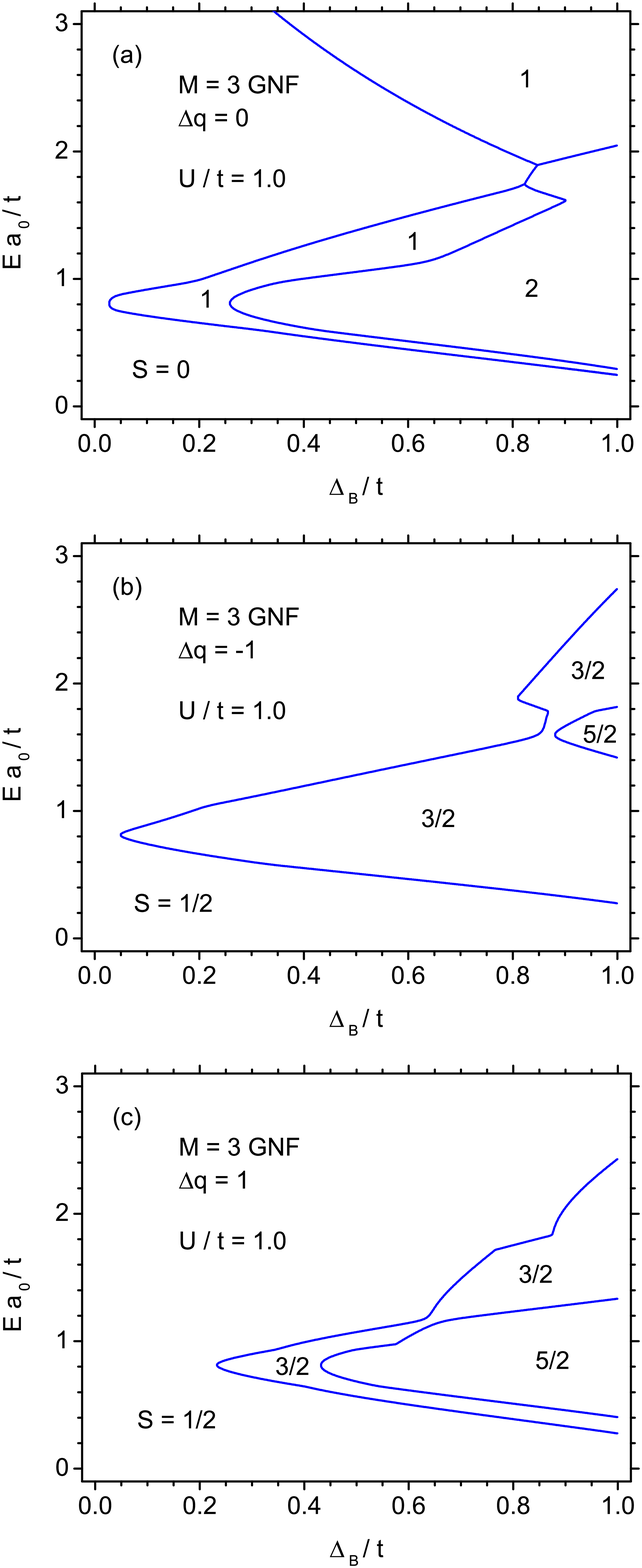}
  \end{center}
   \caption{\label{fig:fig2} Total spin ground-state phase diagram for triangular nanoflake with armchair edges consisting of $M=3$ hexagons, for charge-neutral case (a), doping with a single hole (b) and doping with a single electron (c).}
\end{figure}

\section{Results}

We commence the discussion of the numerical results obtained within the model described above by presenting the total spin phase diagram, in which stability ranges of phases with various values of ($z$ component of) total spin $S$ are determined as a function of normalized external electric field $Ea_0/t$ and normalized external magnetic field $\Delta_B/t$. 

Let us now analyse the ground state phase diagram for the triangular armchair-edged nanoflakes (depicted schematically in Fig.~\ref{fig:fig1}(b)). We start form the smallest structure, i.e. this with $M=3$ hexagons at each edge. Its phase diagram for charge neutrality ($\Delta q=0$) is presented in Fig.~\ref{fig:fig2} (a), where the borders between phases of different values of total spin are indicated as a function of external electric field and external magnetic field. Let us emphasize that in absence of external fields the nanoflake does not exhibit net magnetic moment, what is in concert with the (second) Lieb theorem \cite{lieb}, since the number of carbon atoms in both sublattices is equal and the nanostructure is not doped with charge (i.e. we deal with half-filling of the available energy states). For low $E$ fields, such a nonmagnetic state is rather persistent and survives up to high magnetic fields. Moreover, below some critical magnetic field (for $\Delta_{B}/t\lesssim 0.028$) also the applied electric field does not enforce any transition to $S>0$ phase. Above the critical $\Delta_{B}$, electric field causes a finite range of state with total spin of $S=1$ to emerge. At moderate electric fields it is soon replaced with $S=2$ when magnetic field increases. However, the range of $S=1$ ordering builds up also at strongest $E$ for $\Delta_{B}$ high enough. In the vicinity of this range we also notice a phase border directly between $S=0$ and $S=2$. 

In order to analyse the effect of weak charge-doping on the phase diagram, we present Fig.~\ref{fig:fig2}(b) and (c), where the same nanostructure is considered, but doped with a single charge carrier - hole ($\Delta q=-1$) in Fig.~\ref{fig:fig2}(b) and electron ($\Delta q=1$) in Fig.~\ref{fig:fig2}(c). First, let us notice that the sign of charge doping is important for the phase diagram, since both plots for $\Delta q=\pm 1$ are different. For both cases of doping, spin $S=1/2$ is predicted for the absence of external fields, due to odd total number of electrons. This low-spin phase persists in the whole studied range of $E$ unless $\Delta_B$ exceeds some critical value, which is significantly higher for electron doping ($\Delta_{B}/t\simeq 0.233$) than for hole doping ($\Delta_{B}/t\simeq 0.05$). For hole doping, a range of spin-$3/2$ ordering emerges at moderate electric field and becomes wider as $\Delta_{B}$ rises. Only a limited area of phase with $S=5/2$ is present for high magnetic field. On the other hand, for electron doping, the ordering with $S=3/2$ appears at higher magnetic fields, and becomes soon replaced with $S=5/2$. The remarkable feature is the disappearance of the higher spin state, which has been found for charge neutral nanostructures at highest electric fields for $\Delta_{B}$ large enough. Therefore, weak charge doping tends to favour the low-spin phase.

\begin{figure}[h!]
  \begin{center}
\includegraphics[scale=0.25]{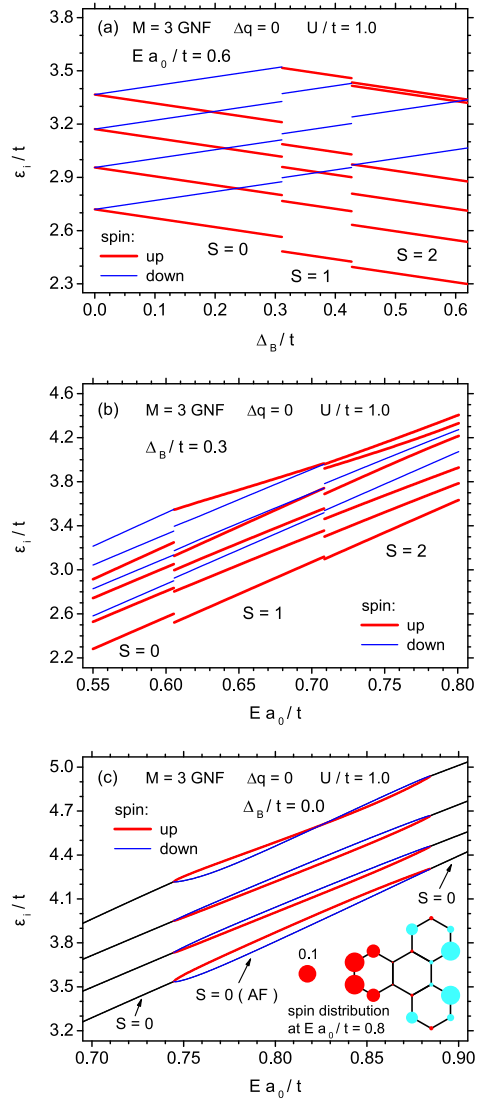}
  \end{center}
   \caption{\label{fig:fig3} Energies of a few single-electron occupied states of highest energy for a triangular nanoflake with armchair edges with $M=3$. Energy is plotted: (a) for constant electric field $Ea_0/t=0.6$ as a function of magnetic field $\Delta_{B}/t$; (b) for constant magnetic field $\Delta_{B}/t=0.3$ as a function of electric field $Ea_0/t$; (c) for constant magnetic field $\Delta_{B}/t=0.0$ as a function of electric field $Ea_0/t$. The inset to (c) shows spin density distribution at $Ea_0/t=0.8$.}
\end{figure}

\begin{figure}[h!]
  \begin{center}
   \includegraphics[scale=0.25]{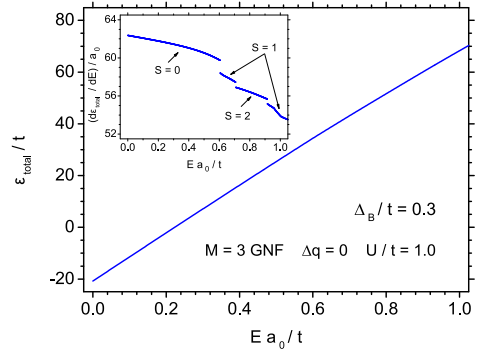}
  \end{center}
   \caption{\label{fig:fignew} Normalized total energy of charge carriers as a function of external normalized electric field for a triangular nanoflake with armchair edges with $M=3$, for constant magnetic field $\Delta_{B}/t=0.3$. Inset presents the analogous dependence of normalized derivative of total energy with respect to electric field.}
\end{figure}

In order to explain the mechanism of the observed magnetic polarization under the influence of external magnetic or electric field, let us refer to the behaviour of the single particle energy states. In Fig.~\ref{fig:fig3} we present the evolution of the single electron energies (for occupied states only) for a charge-neutral triangular nanoflake with armchair edges with $M=3$ (the phase diagram of which has been presented in the previous figure). Let us remind that in our self-consistent procedure, the energies of single electron states depend on the total charge distribution and on occupation of all the states, i.e. on the total number of electrons in the system. In Fig.~\ref{fig:fig3}(a) we analyse the behaviour of energies as a function of magnetic field for constant electric field of $Ea_0/t=0.6$. Only a few occupied states with highest energies are plotted, including the highest occupied molecular orbital (but equal number of spin-up and spin-down states is omitted). For fields below $\Delta_{B}/t\simeq 0.31$, the nanoflake is in non-magnetic state with $S=0$, so that equal number of electrons occupy spin-up and spin-down states. Let us observe that due to presence of nonzero magnetic field no spin degeneracy takes place and the energies of spin-up and spin-down states are splitted (the splitting varies quasi-linearly with the field and states with opposite spin show opposite slopes). The curves for spin-up and spin-down states have common origin at zero magnetic field, since without field the states are spin-degenerate. For the fields in the range of $0.31\lesssim\Delta_{B}/t\lesssim 0.43$ the state with $S=1$ emerges. This is microscopically reflected in the fact that it is more energetically favourable for one electron of spin-down to skip to another state with spin up, so total spin rises by one. This happens to the electron with highest energy among those with spin-down (when $S=0$). When the state $S=0$ turns into $S>0$ a discontinuous change in energy splitting of the single electrons states takes place, as the total nonzero magnetization appears in the system. The total decrease in energy for spin-up electrons exceeds the rise in energy for spin-down electrons. When $\Delta_{B}/t$ reaches the value of approximately 0.43, further transition from $S=1$ to $S=2$ occurs. Its mechanism is analogous to the previous transition; again, the spin-down electron with highest energy skips to the state with spin-up. Therefore, the observed kind of behaviour is owing to single energy states and their behaviour in magnetic field. Let us remind that our system of interest is molecular-like and possesses a discrete set of energy states.

Fig.~\ref{fig:fig3}(b) presents the dependence of single particle energy states on electric field at constant magnetic field $\Delta_{B}/t=0.3$. Here, energies of both spin-up and spin-down states indicate the same sign of slope in the field. This time the Zeeman splitting does not depend on $E$ field unless total spin is changed, since the magnetic field is constant. In a way similar to the previous case, transitions between $S=0$ and $S=1$ (at $Ea_0/t\simeq 0.61$) or between $S=1$ and $S=2$ (at $Ea_0/t\simeq 0.91$) consist in skipping of single electrons from spin-down to spin-up states. 

In Fig.~\ref{fig:fig3}(c) we plot again the dependence of single particle energy states on electric field, but this time at zero magnetic field. Below the field of approximately $Ea_0/t\simeq 0.74$, the states are spin-degenerate and the total spin of the nanostructure is equal to zero. However, in the range of $0.75\lesssim Ea_0/t\lesssim 0.88$, the spin degeneracy is lifted and Zeeman splitting arises, even though the total spin remains zero. This phenomenon is related with emergence of antiferromagnetic polarization of the nanoflake, as it can be seen in the inset in Fig.~\ref{fig:fig3}(c), where spin distribution is shown for $Ea_0/t=0.80$. As it can be noticed, this kind of antiferromagnetic polarization breaks the rotational symmetry of the nanoflake. It is seen that the Zeeman splitting of the energy states is a continuous function of the field (so that the antiferromagnetic polarization emerges continuously, contrary to ferromagnetic phase). Let us remark that the appearance of antiferromagnetic states for the nanoflakes being short sections of armchair nanoribbon with zigzag terminations will be discussed in details in further part of this section. 

We mention that a similar phenomenon of appearance of magnetic polarization in a nanoflake under the influence of external electric field has been discussed by us in the Ref.~\cite{Szalowski2014} in the context of controlling the indirect magnetic coupling mediated by the graphene nanostructure (see Fig.~5 in Ref.~\cite{Szalowski2014}).

In Fig.~\ref{fig:fig3} it has been shown that single-particle energies exhibit discontinuous behaviour as a function of field, since they vary in piecewise-continuous and quasi-linear way. However, this is not the case for the total energy of the systems of charge carriers, which remains a strictly continuous function of external electric and magnetic field. Such a property is illustrated in Fig.~\ref{fig:fignew}, in which total energy is plotted as a function of electric field for a charge-neutral triangular nanoflake with armchair edges with $M=3$ in magnetic field of $\Delta_{B}/t=0.3$. The dependence is quasi-linear with slowly decreasing slope. This slope, i.e. the derivative of total energy with respect to electric field is shown in the inset in Fig.~\ref{fig:fignew}. For this quantity discontinuous behaviour is visible, with abrupt changes of derivative value at certain critical electric fields, which correspond to transitions between states with different total spin (the value of which is indicated in the plot). This behaviour can be compared with the phase diagram shown in Fig.~\ref{fig:fig2}(a) and it is also consistent with Fig.~\ref{fig:fig3}(b). Such an observation confirms that we deal with discontinuous, first-order phase transitions. On the other hand, the dependence of total energy on electric field has been also studied for the case when antiferromagnetic ordering emerges (as shown in Fig.~\ref{fig:fig3}(c)). In such a case, not only the total energy is continuous, but also its first derivative indicates this feature. It is the second derivative of the total energy which exhibits discontinuities at the critical fields where antiferromagnetic ordering emerges (see Fig.~\ref{fig:fig3}(c)), what indicates that such phase transitions are of second order (continuous).

\begin{figure}[h!]
  \begin{center}
   \includegraphics[scale=0.25]{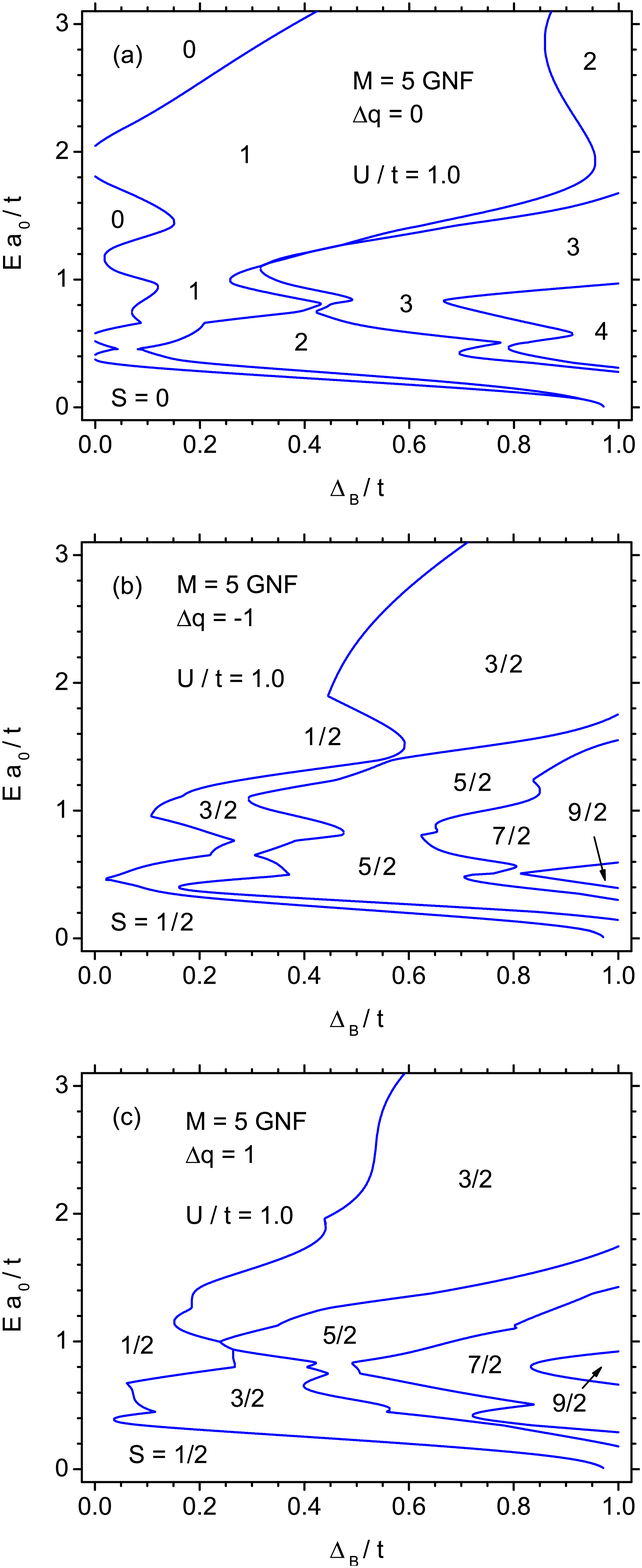}
  \end{center}
   \caption{\label{fig:fig4} Total spin ground-state phase diagram for triangular nanoflake with armchair edges consisting of $M=5$ hexagons, for charge-neutral case (a), doping with a single hole (b) and doping with a single electron (c).}
\end{figure}

Let us now discuss the phase diagram for larger nanoflakes with armchair edges. In Fig.~\ref{fig:fig4} (a) we present the ground state phase diagram for a charge-undoped armchair triangular nanoflake with $M=5$ hexagons at the edge. Like for $M=3$, the state with $S=0$ remains robust against magnetic field for low electric field. However, at zero (or low enough) magnetic field, application of electric field causes a series of transitions between states with $S=0$ and $S=1$ (with relatively low critical values of $E$). If $\Delta_{B}/t$ exceeds approximately 0.15, the phase with $S=0$ vanishes in the intermediate range of electric fields, remaining only at very low or at high $E$. At the same time states with higher spins ($S=2$ and $S=3$) appear, in particular for low but non-zero $E$. For $\Delta_B/t\gtrsim 0.67$, also an electric field-induced state with spin $S=4$ appears. The phase diagram is richer in comparison with the case of smaller nanostructure with $M=3$; moreover, the electric field-driven magnetic polarization at zero magnetic field is possible for $M=5$. Also states with higher spin appear , in general, for lower values of $\Delta_{B}$ then previously and the total area of nonmagnetic phase is reduced. On the contrary, the area of stability of $S=1$ phase is greatly expanded.

In order to analyse the effect of weak charge-doping on the phase diagram, we present Fig.~\ref{fig:fig4}(b) and (c), where the same nanostructure is considered, but doped with a single charge carrier (hole or electron). For both signs of doping, the transitions from low-spin to higher spin states become possible only if $\Delta_{B}$ exceeds some low critical value, so in complete absence of magnetic field $E$ cannot increase the magnetic moment. From comparison of Fig.~\ref{fig:fig4}(b) and (c) it is seen that for electron doping (Fig.~\ref{fig:fig4}(c)), the state with $S=3/2$ is in general more stable (occupies larger area of the phase diagram) that for hole doping (Fig.~\ref{fig:fig4}(b)). Let us observe that for $S=5/2$ the situation is quite the opposite. However, both phase diagrams bear more qualitative similarity than it was noticeable for smaller nanoflake with $M=3$.

\begin{figure}[h!]
  \begin{center}
\includegraphics[scale=0.25]{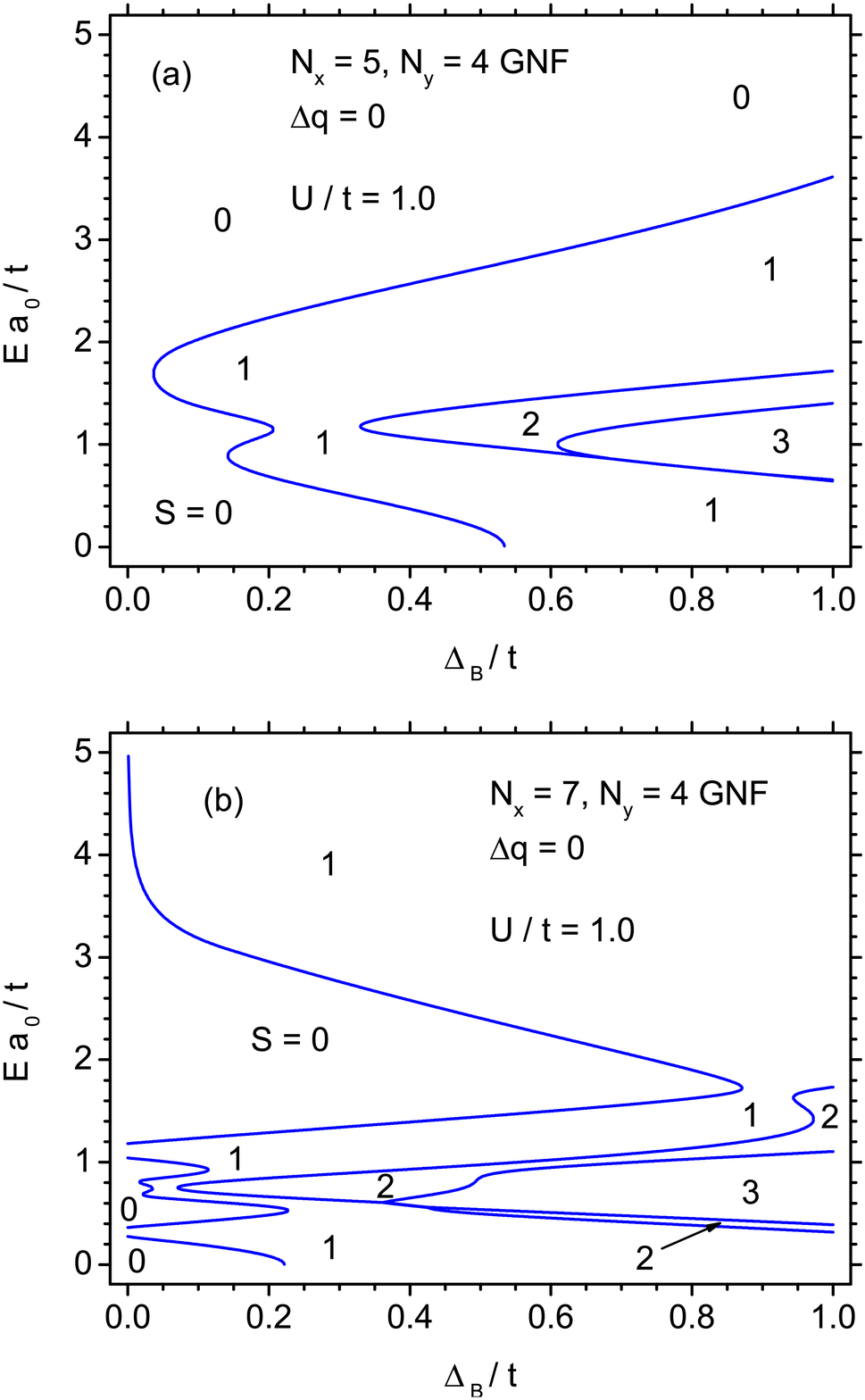}
  \end{center}
   \caption{\label{fig:fig5} Total spin ground-state phase diagram for a short piece of armchair graphene nanoribbon with zigzag terminations, for (a) $N_x=5,N_y=4$ and (b) $N_x=7,N_y=4$. The structures are charge-undoped.}
\end{figure}

A similar ground-state phase diagram as for triangular nanoflakes is also calculated for the case of undoped nanostructures in a form of an ultrashort piece of armchair graphene nanoribbon with zigzag terminations (see Fig.~\ref{fig:fig1}(c)). The diagram for a nanostructure of width $N_x=5$ and length $N_y=4$ is shown in Fig.~\ref{fig:fig5}(a). In the absence of the magnetic field the nanoflake remains in the state of $S=0$ for the whole range of electric fields. Switching on $\Delta_B/t\gtrsim 0.037$ causes the appearance of a finite range of electric fields for which $S=1$ state appears. However, the state with zero total spin is still particularly persistent for strong $E$. If $\Delta_B$ is further increased, also ranges of total spin equal to 2 and then 3 emerge, but only for moderate $E$ values. The diagram is far more complicated for wider nanostructure, with $N_x=7$ and $N_y=4$, as presented in Fig.~\ref{fig:fig5}(b). For zero fields the nanoflake is non-magnetic, but switching on electric field causes the switching to $S=1$ state for some narrow ranges of moderate $E$ for low $\Delta_{B}$, including in particular the zero value. Therefore, spin polarization can emerge in the presence of external electric field and absence of magnetic field. Another, expanding range of $S=1$ phase is noticeable for very strong electric field (unlike the situation found in the narrower nanoflake, i.e. in Fig.~\ref{fig:fig5}(a), where nonmagnetic phase strongly dominates at high $E$). When $\Delta_B/t$ increases above approximately 0.22, the low-field state with zero spin tends to vanish completely, being replaced mainly with phase of $S=1$. Also states with higher spins appear successively at moderate $E$ for stronger magnetic field and this happens for lower $\Delta_{B}$ than in the case of a narrower nanostructure. Again, the diagram for larger nanoflake exhibits more complexity than for the smaller one.

\begin{figure}[h!]
  \begin{center}
\includegraphics[scale=0.25]{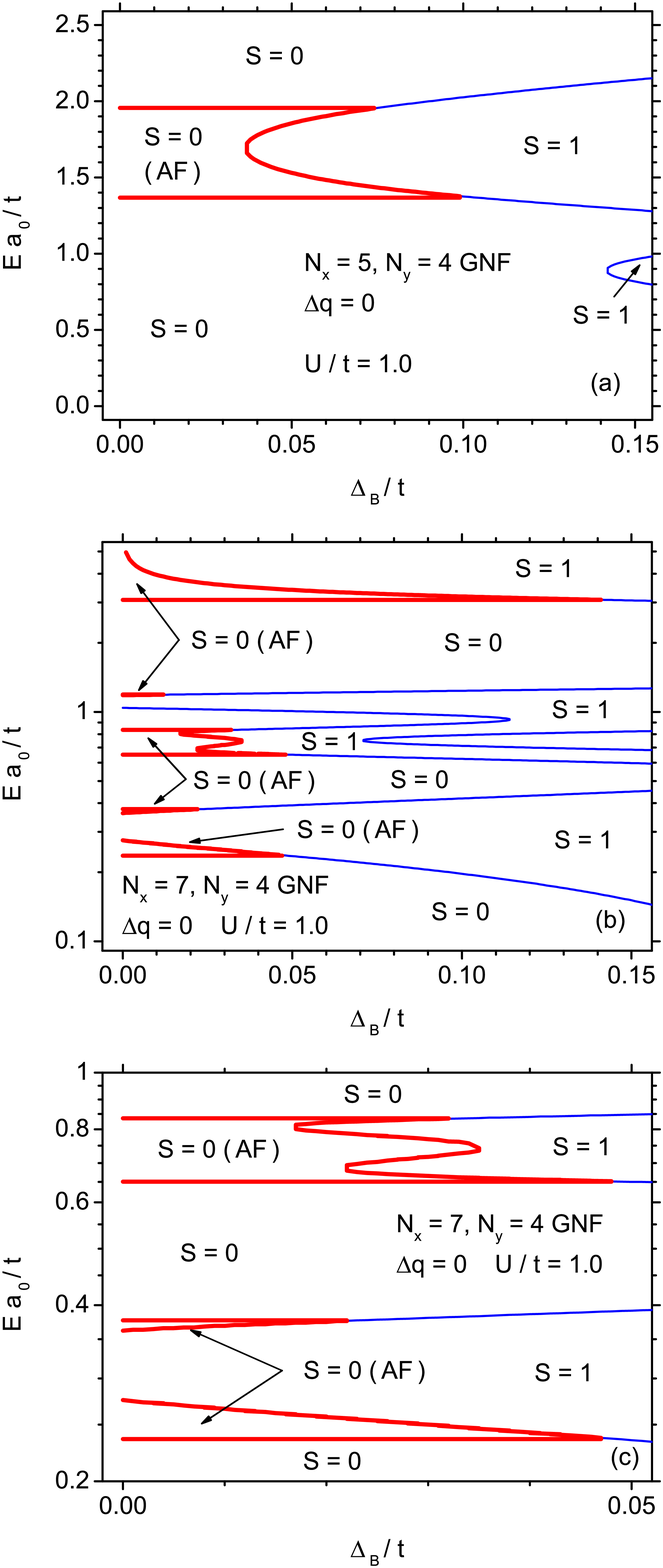}
  \end{center}
   \caption{\label{fig:fig6} Ground-state phase diagram for a short piece of armchair graphene nanoribbon with zigzag terminations with marked stability ranges of antiferromagnetic phase $S=0$(AF); (a) $N_x=5,N_y=4$; (b) $N_x=7,N_y=4$; (c) enlarged plot for $N_x=7,N_y=4$. The structures are charge-undoped.}
\end{figure}

\begin{figure}[h!]
  \begin{center}
\includegraphics[scale=0.25]{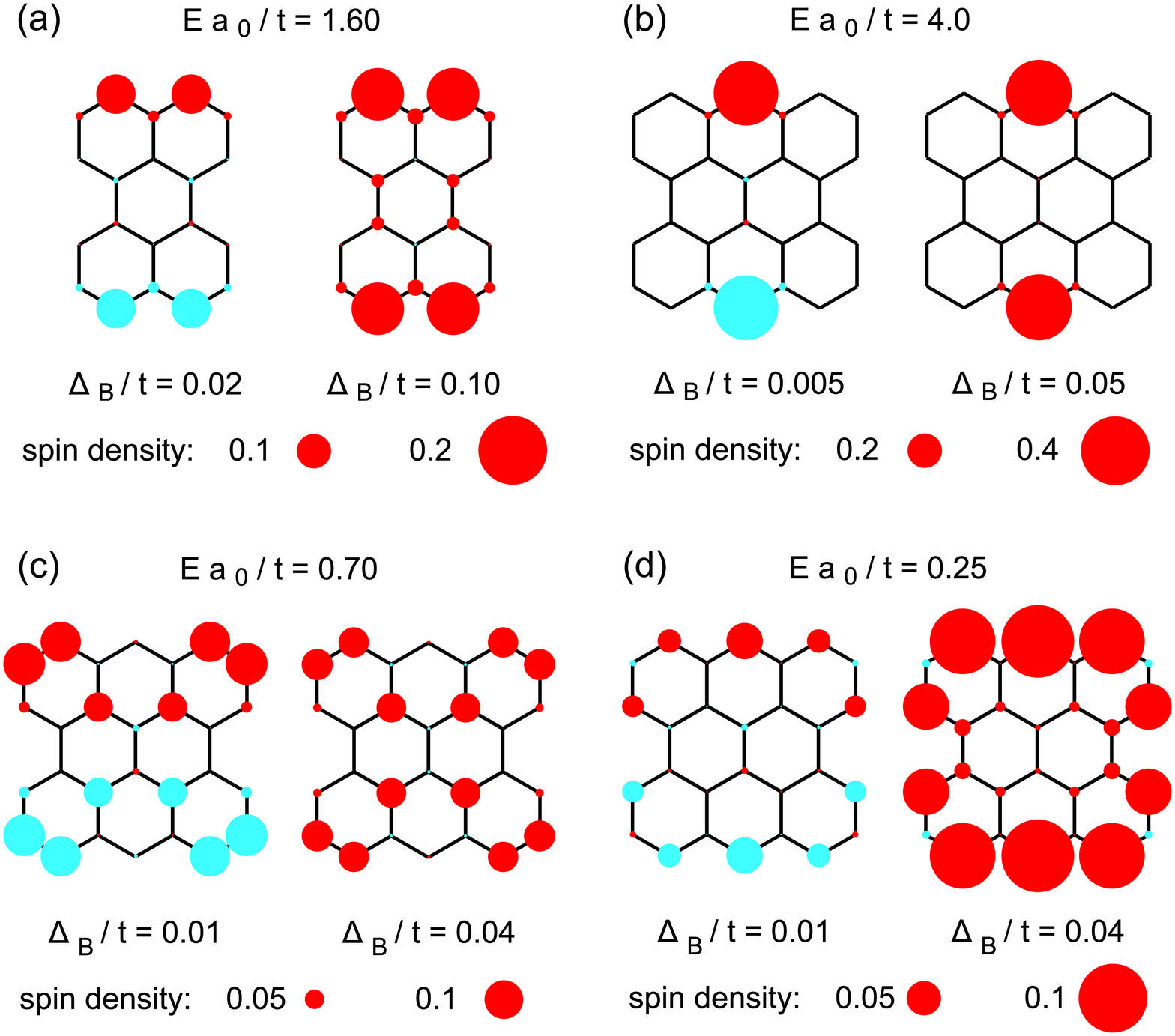}
  \end{center}
   \caption{\label{fig:fig7} Spin density distribution for charge undoped short piece of armchair graphene nanoribbon with zigzag terminations. The nanostructure size and electric field values are: (a) $N_x=5,N_y=4$, $Ea_0/t=1.60$; (b) $N_x=7,N_y=4$, $Ea_0/t=4.0$; (c) $N_x=7,N_y=4$, $Ea_0/t=0.7$; (d) $N_x=7,N_y=4$, $Ea_0/t=0.25$. In each case two values of magnetic field were selected to show antiferromagnetic phase (weaker magnetic field) and ferromagnetic phase (stronger magnetic field). Different colors denote opposite spin orientations, while spin value associated with a given site is proportional to the radius of the circle.}
\end{figure}

Let us emphasize that the zero value of the total spin of the charge-undoped nanostructure does not necessarily mean complete absence of magnetic polarization. In addition to truly nonmagnetic phase, an antiferromagnetic ordering is possible, which consists in opposite magnetic polarizations of both carbon sublattices. Such a possibility has been notices in our earlier work devoted to behaviour of graphene nanostructure-mediated indirect coupling in external electric field (see Ref.~\cite{Szalowski2014}). Let us focus here on this kind of ordering for the case of nanoflakes being short fragments of armchair nanoribbon (see Fig.~\ref{fig:fig1}(c)). In order to separate the phase with $S=0$ into antiferromagetic (AF) one and nonmagnetic one, we present Fig.~\ref{fig:fig6}, where the part of phase diagram for low $\Delta_B$ is enlarged (see the full range diagrams presented in plotted in Fig.~\ref{fig:fig5}(a) and (b)). The stability areas of AF ordering in Fig.~\ref{fig:fig6} are marked with thick solid lines, while the other borders between phases of various total spins are plotted with thinner lines. We mention that in this case the predicted borders of AF phase may correspond to the presence of cross-overs. In Fig.~\ref{fig:fig6}(a) the nanostructure with $N_x=5$ and $N_y=4$ is considered, which exhibits a single range of AF ordering for $\Delta_B\lesssim 0.1$ in a finite range of considerably strong electric field. The spin polarization distribution for two representative points of phase diagram for this nanostructure (AF phase and phase with spin $S=1$) is plotted in Fig.~\ref{fig:fig7}(a). It is visible that spin density is localized mainly at the outermost carbon atoms of the zigzag terminations of the nanoflake. Moreover, the metamagnetic transition enforced by magnetic field at constant $E$ consists not only in reorientation of magnetization direction at one zigzag edge, but also in some increase in absolute value of spin density at edge carbon sites. 

Analogous presence of AF phase ranges is visible for a wider nanostructure ($N_x=7$, $N_y=4$), as shown in Fig.~\ref{fig:fig6}(b) and (c) (and spin density distributions are presented in Fig.~\ref{fig:fig7}). There, a logarithmic scale for electric field has been used to emphasize the behaviour at low $E$ fields. In general, it can be observed that the borders between $S=0$ nonmagnetic and $S=0$ AF phase are always straight horizontal lines, so that only the electric field can cause the sublattice polarization to emerge or disappear. On the other hand, borders separating AF phase and nonzero spin phase exhibit various shapes, so that both magnetic and electric field can cause a metamagnetic transition form antiferromagnetic to ferromagnetically polarized state.

The largest area of AF phase exists in the vicinity of high $E$-field transition between $S=0$ and $S=1$ and this range gradually vanishes if magnetic field increases. In Fig.~\ref{fig:fig7}(b) it is shown that the spin polarization for constant $Ea_0/t=4.0$ is localized again at the armchair edge, practically only at lattice sites at the symmetry axis of the nanoflake. This time metamagnetic transition is not connected with noticeable change in spin density magnitude.

The AF ordering is noticed also at weaker $E$ fields, what is shown in detail in Fig.~\ref{fig:fig7}(c). At $Ea_0/t=0.70$, spin density is concentrated away of the symmetry axis, mainly at zigzag edge close to its ends. Transition from AF to F state as the magnetic field increases is connected with noticeable reduction in spin density values, as seen in Fig.~\ref{fig:fig7}(c)). For the lowest $E$ field at which AF phase exists (Fig.~\ref{fig:fig7}(d)), magnetization is distributed both at zigzag and armchair edges of the structure, with a particularly strong increase in spin density magnitude when the magnetic field causes a metamagnetic transition.

In general it can be noticed that in all the presented cases, the antiferromagnetic ordering is characterized by  spatially well-separated areas of nonzero spin density with opposite direction of spin. 

\begin{figure}[h!]
  \begin{center}
\includegraphics[scale=0.25]{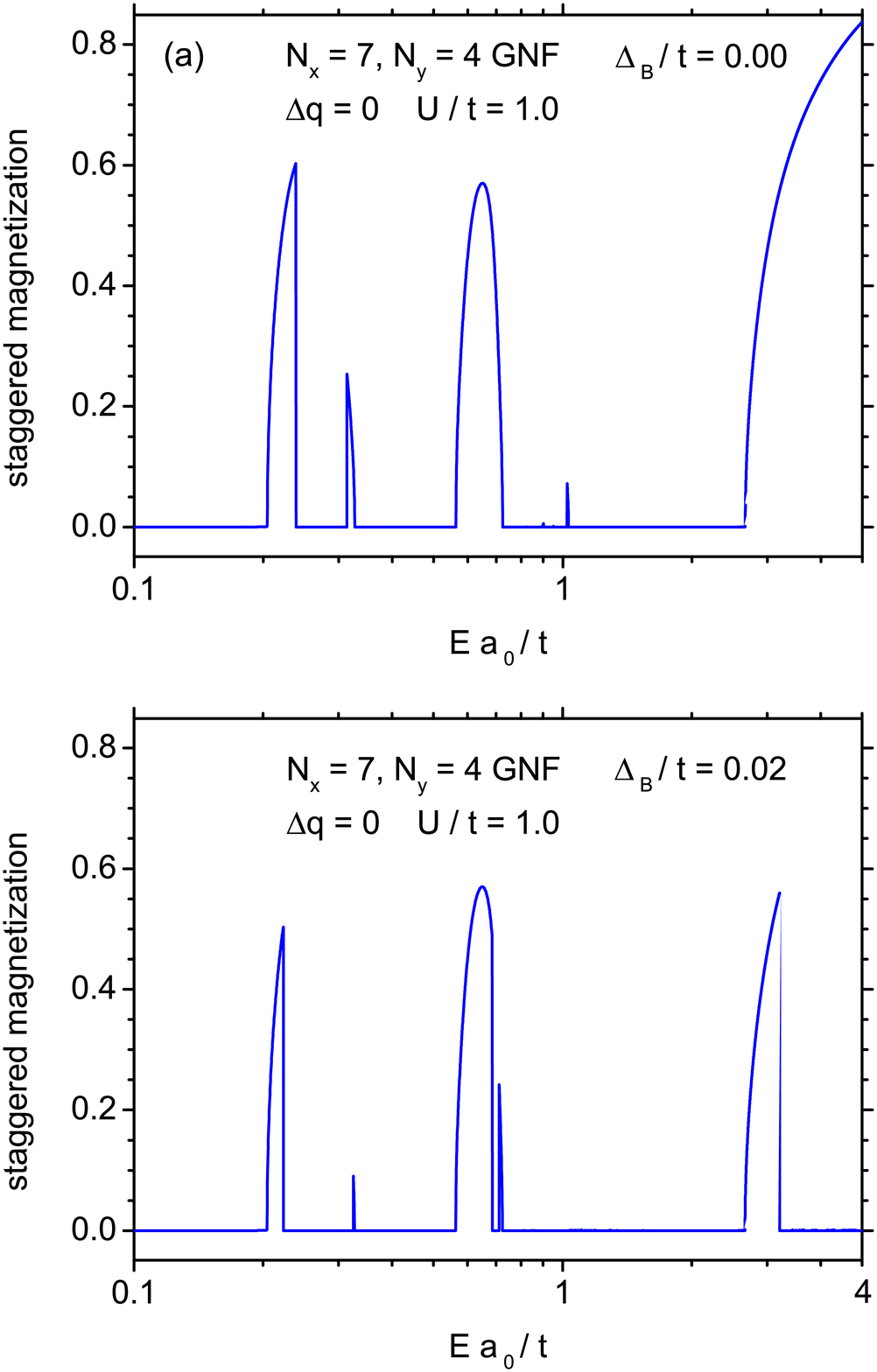}
  \end{center}
   \caption{\label{fig:fig8} Dependence of staggered magnetization on the normalized electric field for undoped short piece of armchair graphene nanoribbon with zigzag terminations ($N_x=5,N_y=4$). Normalized magnetic field is: (a) $\Delta_{B}/t=0.0$ and (b) $\Delta_{B}/t=0.02$.}
\end{figure}

The antiferromagnetic ordering can be characterized with an order parameter called staggered magnetization, which is the difference in magnetization between both carbon sublattices. The evolution of such parameter with changes of electric field $Ea_0/t$ can be followed in Fig.~\ref{fig:fig8} for the absence of magnetic field (Fig.~\ref{fig:fig8}(a)) and in presence of $\Delta_B/t=0.02$ (Fig.~\ref{fig:fig8}(b)). All the ranges of AF phase noticeable in the phase diagram Fig.~\ref{fig:fig6}(b) and (c) can be identified here. It is also visible that the staggered magnetization varies continuously with the electric field if the transition is between $S=0$ nonmagnetic and AF state. On the other hand, transition between $S=0$ AF and $S=1$ state caused by $E$ field are discontinuous (see for example a "window" between $Ea_0/t\simeq 0.25$ and $0.30$, where $S=1$ state appears - such a range is also shown in Fig.~4 of our work \cite{Szalowski2014}). Increase in magnetic field promotes $S=1$ state at the cost of AF $S=0$ ordering and reduces staggered magnetization values especially at high $E$ fields.

\begin{figure}[t]
  \begin{center}
 \includegraphics[scale=0.28]{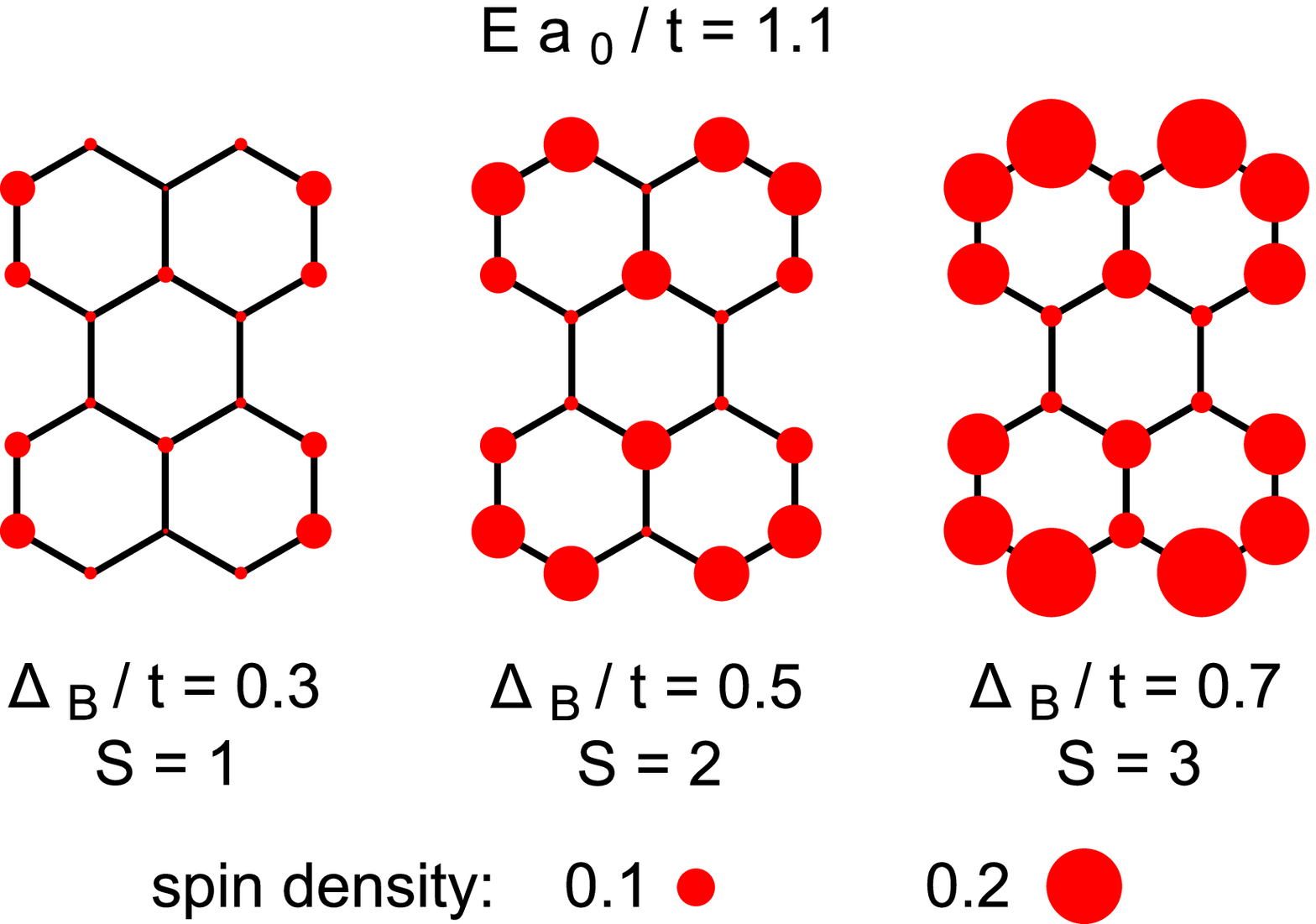} 
  \end{center}
   \caption{\label{fig:fig9} Spin density distribution for charge undoped short piece of armchair graphene nanoribbon with zigzag terminations ($N_x=5,N_y=4$) at constant electric field $Ea_0/t=1.1$, for three values of normalized magnetic field: (a) $\Delta_{B}/t=0.3$ ($S=1$); (b) $\Delta_{B}/t=0.5$ ($S=2$); (c) $\Delta_{B}/t=0.7$ ($S=3$). Spin density associated with a given site is proportional to the radius of the circle.}
\end{figure}

\begin{figure}[t]
  \begin{center}
 \includegraphics[scale=0.28]{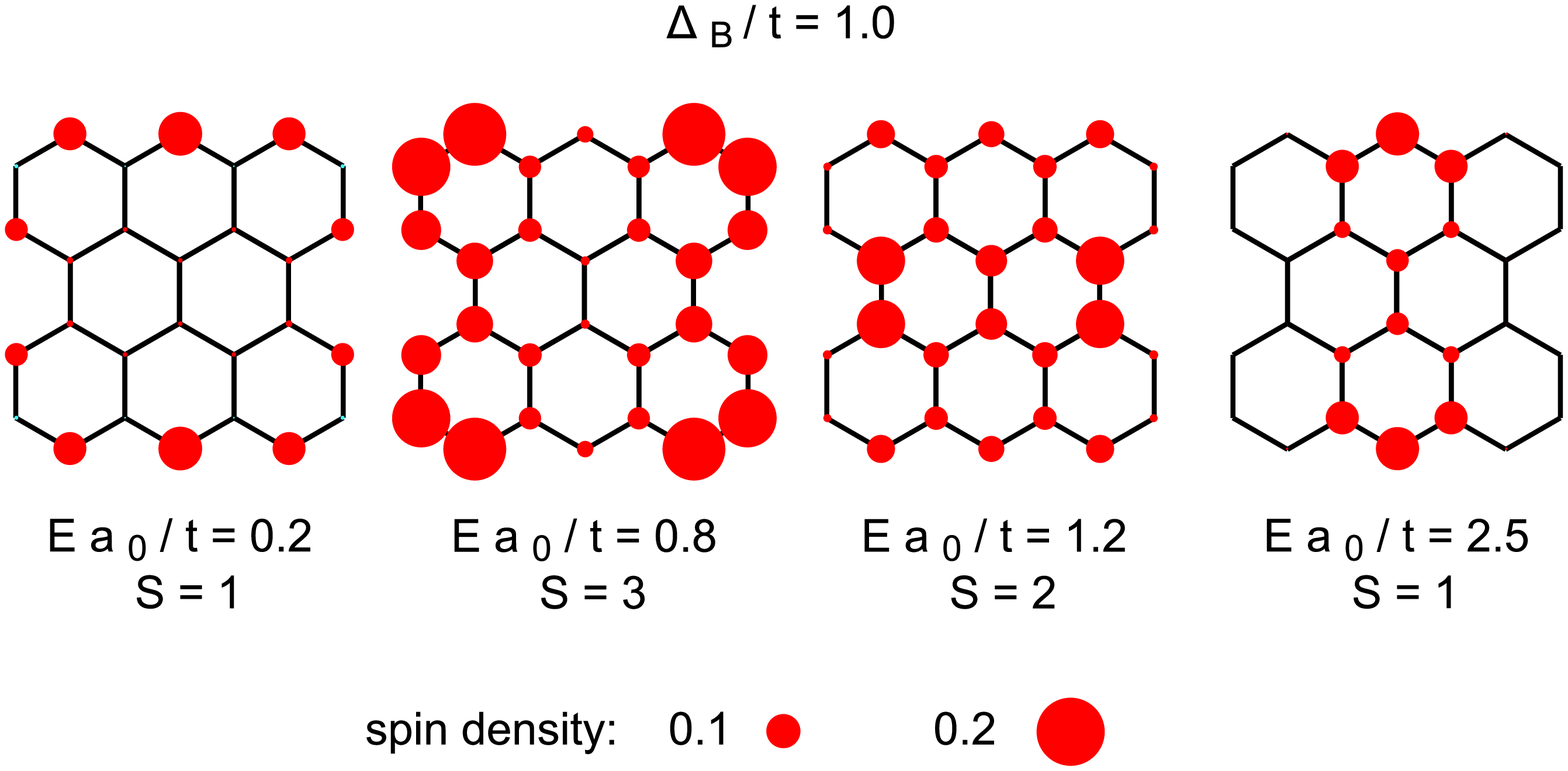}
  \end{center}
   \caption{\label{fig:fig10} Spin density distribution for charge undoped short piece of armchair graphene nanoribbon with zigzag terminations ($N_x=7,N_y=4$) at constant magnetic field $\Delta_{B}/t=1.0$, for three values of normalized electric field: (a) $Ea_0/t=0.2$ ($S=1$); (b) $Ea_0/t=0.8$ ($S=3$); (c) $Ea_0/t=0.2$ ($S=2$); (d) $Ea_0/t=2.5$ ($S=1$). Spin density associated with a given site is proportional to the radius of the circle.}
\end{figure}

\begin{figure}[t]
  \begin{center}
\includegraphics[scale=0.25]{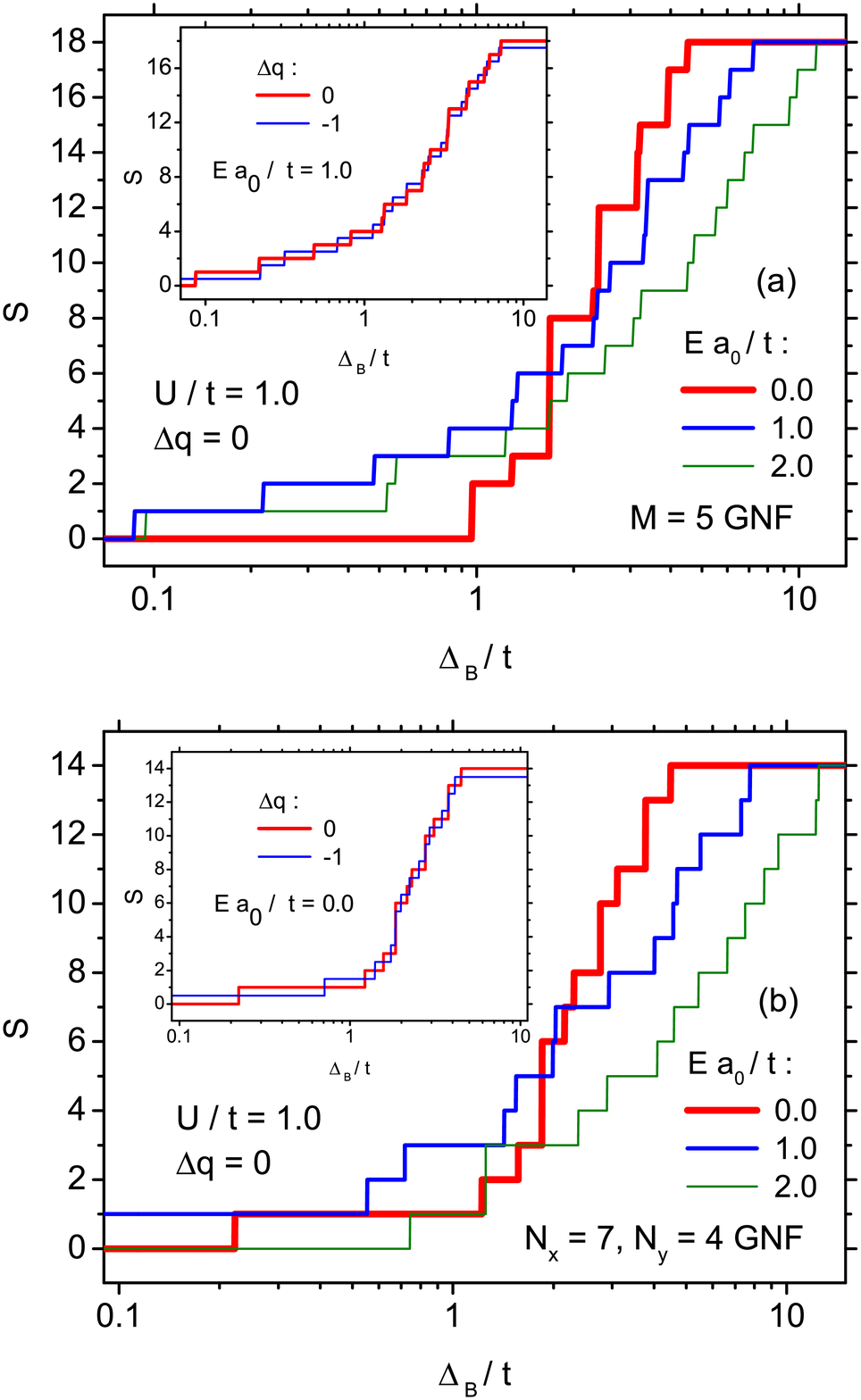}
  \end{center}
   \caption{\label{fig:fig11} Dependence of the total spin of nanostructure on the normalized magnetic field for three values of normalized electric field $Ea_0/t=0.0,1.0$ and $2.0$. (a) triangular nanoflake with armchair edges consisting of $M=5$ hexagons (in the inset the total spin is plotted as a function of magnetic field in the electric field $Ea_0/t=1.0$ for charge-undoped case and for doping with a single hole); (b) short piece of armchair graphene nanoribbon with zigzag terminations, $N_x=7,N_y=4$ (in the inset the total spin is plotted as a function of magnetic field in absence of electric field for charge-undoped case and for doping with a single hole).}
\end{figure}

The process of magnetization of a nanoflake by increasing the magnetic field and related spin distribution can be followed in Fig.~\ref{fig:fig9} for structure with $N_x=5$ and $N_y=4$ (for states with $S=1,2,3$), at fixed electric field $Ea_0/t=1.1$. Magnetization first appears at armchair edges of the nanoflake and then it gradually becomes distributed on both zigzag and armchair edges, dominantly at the outermost carbon atoms. 

A similar process of magnetization distribution changes caused by varying electric field at constant $\Delta_B/t=1.0$ is depicted in Fig.~\ref{fig:fig10} for a nanoflake with $N_x=7$ and $N_y=4$. For weak field $S=1$, magnetization appears mainly at the zigzag edge. Then, for $S=3$ it is strongly pushed to the armchair edges, while further increase in $E$ reduces $S$ to the value of 2 with polarization localized in the middle of the structure. Finally, for the strongest field, again we observe $S=1$, this time with spin density at zigzag terminations but only close to the symmetry axis.

It is also interesting to follow the magnetization process of selected nanoflakes during application of increasing magnetic field at constant electric field, i.e. the dependence of total spin on $\Delta_B/t$. Such calculations are presented in Fig.~\ref{fig:fig11}(a) for a triangular graphene nanoflake with armchair edges composed of $M=5$ hexagons. Three curves in main plot correspond to three values of external electric field and logarithmic scale is used for $\Delta_B/t$ in order to emphasize the low-field behaviour. It is visible that at zero $E$ field, the nanostructure remains in nonmagnetic state up to approximately $\Delta_{B}/t\simeq 1$, while further increase of the field causes a sequence of metamagnetic transitions, leading quite quickly to saturation. Such a behaviour has been also predicted in one of our earlier works \cite{Szalowski2013c} for bowtie-shaped nanostructure (without electric field). Switching on electric field (cases of $Ea_0/t=1.0$ and $2.0$) causes two effects. First, the magnetization process at low magnetic fields becomes faster, i.e. the transitions to higher spin states start at lower $\Delta_B$. Secondly, at higher magnetic fields (what means approximately $\Delta_B/t\gtrsim 2$), the magnetization progress is slower, so that higher field is necessary to reach the saturation. Therefore, in presence of external electric field the shape of dependence of total spin on magnetic field is changed significantly. For low field, $Ea_0/t=1$ causes the magnetic polarization of the flake at zero magnetic field, as discussed before. In order to assess the importance of charge doping we plot an analogous dependence in the inset of Fig.~\ref{fig:fig11}(a) for the same nanostructure doped with a single hole ($\Delta q=-1$) and undoped ($\Delta q=0$) in presence of electric field $Ea_0/t=1.0$. It is visible that the effect of doping on the magnetization process is, in general, not significant. However, let us mention that the state at zero magnetic field corresponds then to $S=1/2$ and also the saturation magnetization is reduced with respect to the undoped case.

Fig.~\ref{fig:fig11}(b) presents similar calculations for an undoped short section of armchair nanoribbon with zigzag terminations ($N_x=7$, $N_y=4$). The qualitative shape of the magnetization curve and the effect of electric field on it bears much similarity to the previous case shown in Fig.~\ref{fig:fig11}(a). In particular, $E$ field causes faster increase in total spin for low magnetic fields and, on the other hand, delays the saturation. In the inset to this figure, magnetization curves for the structure doped with a single hole ($\Delta q=-1$) is compared with the case of charge neutrality in absence of electric field. Again, only a slight influence of charge doping is found.

\section{Final remarks}

In the paper we have studied the mean field ground-state phase diagram for two classes of graphene nanostructures with armchair edges. The selected shapes were: triangular nanoflakes with armchair edges and short fragments of armchair graphene nanoribbons with zigzag terminations. The study was aimed at characterizing the influence of in-plane external electric and magnetic field on the total spin as well as the spin distribution in the nanoflakes. The total energy of the system of the charge carriers was calculated in a self-consistent way for the tight-binding model Hamiltonian supplemented with Hubbard term in MFA as well as terms accounting for external fields. 

Due to the edge form and equal number of carbon atoms in both sublattices, the charge-neutral structures do not exhibit nonzero spin polarization in absence of external fields. However, both electric and magnetic field has been found in general to enable switching to states with nonzero total spin. We noticed that the complexity level of the phase diagram increases significantly with the size of the studied nanostructure (for both shapes of interest in the present study). The states with $S=1$ appear dominantly for moderate electric fields (within the studied range) for weaker magnetic fields and the increase in magnetic field caused further states with $S>1$ to emerge. However, higher electric fields cause rather nonmagnetic state (or state with $S=1$ at higher magnetic fields) to become the the most energetically favourable. Let us observe that a tendency to demagnetization under the action of electric field has been noticed for example in Ref.~\cite{Zhou2013} for bowtie-shaped nanoflakes, in Ref.~\cite{Ma2012} for triangular ones or in Ref.~\cite{Farghadan2014} for various nanorings. The magnetic polarization can emerge at zero magnetic and critical electric field for larger studied structures, while for smaller ones some critical magnetic field is necessary to enable emergence of spin via electric field.

The predicted metamagnetic transitions in the investigated nanostructures emerge as a result of behaviour of single, discrete energy states in external field. 

The influence of weak charge doping on the magnetic phase diagram has been also illustrated for the case of triangular nanoflakes with armchair edge. Significant differences between electron- and hole-doping (with a single charge carrier) were observed, in particular for the smallest nanoflake studied (hole doping promoted the states with higher spins for wider range of fields than electron doping). 
 
For nanoflakes being short pieces of armchair graphene nanoribbon, the phase diagram was enriched by finding the antiferromagnetic phases within the range of $S=0$. All such phases, stable only in weak magnetic field and nonzero electric field, were characterized by good spatial separation of spin density ranges with opposite spin orientation. Moreover, the metamagnetic transitions between $S=0$ antiferromagnetic and non-magnetic state are only possible by varying the electric field, not the magnetic field (which can enforce the state with $S>0$).

The process of magnetization of the nanoflake with magnetic field at constant electric field has been also characterized. Electric field has been found to shift the magnetic saturation towards higher magnetic fields, while, on the other hand, it promoted faster magnetization at weaker magnetic fields.

We are convinced that our work can inspire further research focused on exploring the phase diagrams of graphene nanostructures in both electric and magnetic field, for which the MFA results provide a reference point. In particular, since the magnetic field can originate from the ferromagnetic substrate, it can be of importance for prediction of behaviour of graphene nanostructures placed on such substrate in external electric field.

\section{Acknowledgments}

\noindent The computational support on Hugo cluster at Laboratory of Theoretical Aspects of Quantum Magnetism and Statistical Physics, P. J. \v{S}af\'{a}rik University in Ko\v{s}ice is gratefully acknowledged.

\noindent This work has been supported by Polish Ministry of Science and Higher Education on a special purpose grant to fund the research and development activities and tasks associated with them, serving the development of young
scientists and doctoral students.

\vspace*{0.5cm}

\section*{References}

\bibliographystyle{elsarticle-num}

\end{document}